\documentclass{JHEP3}
\usepackage{epsfig,multicol,bbm}
\newcommand{\pio}{\mbox{$\pi^0$}}

\newcommand{\ks}{\mbox{$K_S$}}
\newcommand{\kl}{\mbox{$K_L$}}

\def\ifm#1{\relax\ifmmode#1\else$#1$\fi}
\def\to{\ifm{\rightarrow}}
\title{\mathversion{bold}
Measurement of the  $K_S\rightarrow \gamma \gamma$
branching ratio using a pure $\ks$ beam with
the KLOE detector}

\author{The KLOE Collaboration:\\
F.~Ambrosino$^{c,d}$,
A.~Antonelli$^a$,
M.~Antonelli$^a$,
F.~Archilli$^a$,
P.~Beltrame$^b$,
G.~Bencivenni$^a$,
S.~Bertolucci$^a$,
C.~Bini$^{f,g}$,
C.~Bloise$^a$,
S.~Bocchetta$^{h,i}$,
F.~Bossi$^a$,
P.~Branchini$^i$,
P.~Campana$^a$,
G.~Capon$^a$,
T.~Capussela$^a$,
F.~Ceradini$^{h,i}$,
F.~Cesario$^{h,i}$,
P.~Ciambrone$^a$,
F.~Crucianelli$^f$,
S.~Conetti$^l$,
E.~De~Lucia$^a$,
A.~De~Santis$^{f,g}$,
P.~De~Simone$^a$,
G.~De~Zorzi$^{f,g}$,
A.~Denig$^b$,
A.~Di~Domenico$^{f,g}$,
C.~Di~Donato$^{d}$,
B.~Di~Micco$^{h,i}$,
M.~Dreucci$^a$,
G.~Felici$^a$,
M.~L.~Ferrer$^a$,
S.~Fiore$^{f,g}$,
P.~Franzini$^{f,g}$,
C.~Gatti$^a$,
P.~Gauzzi$^{f,g}$,
S.~Giovannella$^a$,
E.~Graziani$^i$,
W.~Kluge$^b$,
V.~Kulikov$^k$
G.~Lanfranchi$^a$,
J.~Lee-Franzini$^{a,j}$,
D.~Leone$^b$,
M.~Martini$^{a}$\footnote{Corresponding author.
Email address: matteo.martini@lnf.infn.it},
P.~Massarotti$^{c,d}$,
S.~Meola$^{c,d}$,
S.~Miscetti$^{a}$\footnote{Corresponding author.
Email address: stefano.miscetti@lnf.infn.it},
M.~Moulson$^a$,
S.~M\"uller$^a$,
F.~Murtas$^a$,
M.~Napolitano$^{c,d}$,
F.~Nguyen$^{h,i}$,
M.~Palutan$^a$,
E.~Pasqualucci$^g$,
A.~Passeri$^i$,
V.~Patera$^{a,e}$,
F.~Perfetto$^{c,d}$,
P.~Santangelo$^a$,
B.~Sciascia$^a$,
A.~Sciubba$^{a,e}$,
A.~Sibidanov$^{a}$,
T.~Spadaro$^a$,
M.~Testa$^{f,g}$,
L.~Tortora$^i$,
P.~Valente$^g$,
G.~Venanzoni$^a$,
R.~Versaci$^a$.
\\
\setcounter{page}{0}
\thispagestyle{empty}
\\
\llap{$^a$}Laboratori Nazionali di Frascati dell'INFN, Frascati, Italy\\
\llap{$^b$}Institut f\"ur Experimentelle Kernphysik, Universit\"at
Karlsruhe, Germany\\
\llap{$^c$}Dipartimento di Scienze Fisiche dell'Universit\`a
``Federico II'', Napoli, Italy\\
\llap{$^d$}INFN Sezione di Napoli, Napoli, Italy\\
\llap{$^e$}Dipartimento di Energetica dell'Universit\`a
``La Sapienza'', Roma, Italy \\
\llap{$^f$}Dipartimento di Fisica dell'Universit\`a
``La Sapienza'', Roma, Italy\\
\llap{$^g$}INFN Sezione di Roma, Roma, Italy\\
\llap{$^h$}Dipartimento di Fisica dell'Universit\`a
``Roma Tre'', Roma, Italy\\
\llap{$^i$}INFN Sezione di Roma Tre, Roma, Italy\\
\llap{$^j$}Physics Department, State University of New York at
Stony Brook, USA\\
\llap{$^k$}Institute for Theoretical and Experimental
Physics, Moscow, Russia.\\
\llap{$^l$}Physics Department, University of Virginia, McCormick rd,
PO Box 400714, Charlottesville, 22904, Virginia, USA. \\

}

\preprint{\ldots}

  \def\DAF{DA\char8NE}
\def\ifm#1{\relax\ifmmode#1\else$#1$\fi}
\def\figb#1;#2;{\parbox{#2cm}{\epsfig{file=#1.eps,width=#2cm}}}  \def\epm{\ifm{e^+e^-}}
\def\figbc#1;#2;{\cl{\figb #1;#2;}}   
\def\ie{{\it\kern-1pt i.\kern-.5pt e.\kern-.2pt}}  
\let\cl=\centerline  \def\gam{\ifm{\gamma}}  \def\ggam{\gam\gam} \def\x{\ifm{\times}}  
\def\up#1{\ifm{^{#1}}}  \def\dn#1{\ifm{_{#1}}}  \def\ab{\ifm{\sim}}    
\def\L{\ifm{{\cal L}}}  \def\pt#1,#2,{\ifm{#1\x10^{#2}}}  \def\dt{\ifm{{\rm d}\,t}}
\def\pic{\ifm{\pi^+\pi^-}}   \def\minus{$-$}  \def\ord#1;{\ifm{{\mathcal O}(#1)}}
   
\def\bye{\end{document}}

\abstract{
We have searched for the decay $K_S \to \gamma\gamma$
in a sample of $\sim 2 \times 10^9$ $\phi \to K_S K_L$
decays collected at DA$\Phi$NE with an integrated
luminosity of 1.9 fb$^{-1}$.
$K_S$ are tagged by the $K_L$ interaction in the calorimeter.
Two prompt photons must also be detected.
Kinematic constraints reduce the initial 6$\times 10^5$ events to 2740
candidates, from which a signal of 711 $\pm$ 35 events is extracted.
By normalizing to the $\ks\ \to 2 \pi^0$ decays
counted in the same sample, we measure
${\rm BR} (\ks \to \gamma\gamma) =
(2.26 \pm 0.12_{\rm stat} \pm 0.06_{\rm syst} ) \times 10^{-6}$,
in agreement with ${\mathcal O}(p^4)$ Chiral Perturbation Theory predictions.
}

\keywords{\epm\ Experiments}

\begin{document}

\section{Introduction}
A precise measurement of the \ks\to\gam\gam\ partial width provides a test of Chiral Perturbation Theory ($\chi$PT).
The \ks\to2\gam\ decay amplitude has been evaluated at leading order in $\chi$PT,  $\mathcal O(p^4)$,
providing an estimate to a few percent  accuracy of branching ratio (BR):
BR(\ks\to2\gam)= \pt2.1,-6, \cite{ambrosio}.
Measurements of such BR have changed considerably with time~\cite{na31,na48-1} while improving in precision.
The latest determination comes from NA48 \cite{na48-2}, BR=\pt(2.71\pm0.07),-6,.
This result differs by about 30\% from the $\mathcal O(p^4)$ $\chi$PT estimate, possibly due to higher order corrections.

We report in the following on a measurement based on a integrated luminosity $\int\!\L\dt\,$\ab\break1.9 fb\up{-1} collected
with the KLOE detector~\cite{kloe_all} at \DAF\ \cite{dafne}, the Frascati $\phi$-factory.
DA\char8NE is an $e^+e^-$ collider operated at a center of mass energy, $W$, of $\sim 1020$~MeV,
the mass of the $\phi$-meson. Equal-energy positron and electron beams collide at an angle of ($\pi$-0.025) radians,
producing $\phi$-mesons nearly at rest. $\phi$-mesons decay 34\% of the time into nearly collinear $K^0\overline{\hbox{$K^0$}}$ pairs.
Since $J^{PC}(\phi)=1^{--}$, the $K^0\overline{\hbox{$K^0$}}$ pair is in an antisymmetric state and the two kaons are always a pure \ks\kl\ pair.
Detection of a \kl-meson therefore guarantees the presence of a \ks-meson of known momentum
and direction. This procedure, called tagging, allows us to obtain a pure
\ks\ beam. The data analyzed consists of some 2 billions \ks\kl\  pairs.

\section{The KLOE detector}

The KLOE detector consists of a large cylindrical drift
chamber, DC \cite{kloe1}, of 4 m diameter and 3.3 m length
operated with a low $Z$ and density gas (helium-isoC\dn4H\dn10), surrounded by a
lead-scintillating fiber calorimeter,
EMC~\cite{kloe2}. The chamber provides tracking, measuring momenta with a resolution
of $\delta p_\bot/p_\bot$ of 0.4\% at large angle and reconstruction of two track
intersections, vertices, to an accuracy of \ab3 mm. A superconducting coil around
the EMC provides a 0.52 T magnetic field. The low-beta insertion
quadrupoles are in the middle of KLOE. They are therefore
surrounded by two compact tile calorimeters, QCAL \cite{kloe3}, used as veto
for otherwise undetected photons absorbed by the quadrupoles.

The EMC is divided into a barrel and two endcaps covering
98\% of the solid angle. Modules are read out at both ends by
photomultipliers, PM, with a readout granularity of \ab4.4\x4.4 cm\up2
for a total of 2440 cells. The calorimeter thickness is \ab 15 radiation lengths, $X_0$.
Both amplitude and  time information are obtained from the PMs. The signal
amplitude measures the energy deposited in a cell and
its time provides both the arrival time of particles and
the position along the modules of the energy deposits, the latter by time difference.
Cells close in time and space are grouped into a
``calorimeter cluster''.
The cluster energy $E$ is the sum of the cell energies.
The cluster time $T$ and position {\bf R}
are energy-weighted averages.
{\bf R} indicates the cluster position with respect to the
detector origin of coordinates.
Energy and time resolutions are $\sigma_E/E =
5.7\%/\sqrt{E\ {\rm(GeV)}}$ and
$\sigma_t = 57\ {\rm ps}/\sqrt{E\ {\rm(GeV)}}
\oplus100\ {\rm ps}$, respectively.
The photon detection efficiency is $\sim 90\%$ at
$E=20$ MeV and reaches 100\% above 70 MeV.

The QCAL calorimeters, \ab5$X_0$
thick, have a polar angle coverage of 0.94$\,<|\cos\theta|<\break$0.99.
Each calorimeter consists of  16 azimuthal sectors of lead and scintillator tiles.
The readout is by wavelength shifter fibers and photomultipliers. The fiber arrangement allows
the measurement of the longitudinal coordinate by time differences.

Only calorimeter signals are used for the trigger \cite{kloe4}.
Two isolated energy deposits, $E>50$ MeV in the barrel and  $E>150$ MeV in the
endcaps, are required. Identification and rejection of cosmic-ray events
are also performed by the trigger hardware.
A background rejection filter, Filfo~\cite{NIM},
based on calorimeter information runs offline. Filfo rejects residual cosmic-ray,
machine background and Bhabha events degraded by grazing the QCAL, before running event reconstruction.

\section{\mathversion{bold}Search of $K_S\rightarrow\gamma\gamma$ with a pure $\ks$ beam}
\subsection{$K_S$ tagging and event preselection}
\label{presel}
The mean \ks\ and \kl\ decay lengths in KLOE are $\lambda_S \sim 0.6$ cm and $\lambda_L \sim 340$ cm respectively.
About 50\% of the produced \kl-mesons reach the calorimeter before decaying.
\ks-mesons are very cleanly tagged, with high efficiency \ab30\%, by identifying a \kl\ interaction in the EMC, which we call \kl-crash.
A \kl-crash has a very distinctive EMC signature: a late
($\langle\beta_K\rangle\cong0.22)$ high-energy cluster with no nearby track.
The average value of the \epm\ collision center of mass energy, W, is obtained
with an accuracy of 30 keV for each 100 nb$^{-1}$ of integrated luminosity,
by reconstructing large angle Bhabha scattering events.
The mean interaction point, IP, position and the $p_{\phi}$ momentum are
also obtained.
The value of $W$, $p_{\phi}$
and the \kl-crash cluster position provide,
for each event, the trajectory of the $K_S$
with an angular resolution of 1$^{\circ}$ and a momentum
resolution better than 1 MeV.  In the analyzed sample, corresponding to an integrated luminosity
$\int\!\L\dt$=1.9 fb$^{-1}$, we observe \ab\pt700,6, tagged \ks-mesons. Using the most recent value
of BR($\ks\to\gamma\gamma$) \cite{na48-2}, we expect \ab1900
tagged $K_S \to \gamma \gamma$ events.
Because of tagging, we have no \kl\to2\gam\ background,
the major contamination in the NA48 measurement.
The main background in our analysis is due to
$K_S\rightarrow 2\pi^0$ events with two photons undetected
because out of geometrical acceptance or not reconstructed in the EMC.

We estimate all backgrounds with the KLOE Monte Carlo,
MC, \cite{NIM}. We produced $\phi$  decays to all channels
corresponding to an integrated luminosity $\int\!\L\dt\,$\ab1.5 fb$^{-1}$.
In addition, for the signal we use a very large sample of
MC $K_S \to \gamma \gamma$ events, equivalent to $\int\!\L\dt$\ab100 fb$^{-1}$.
In the simulation, the photon detection efficiency and resolutions
have been tuned with data using a large sample of tagged photons
from $\phi \to \pi^+\pi^-\pi^0$  events
selected using only drift chamber information~\cite{NIM}.
\kl\ interactions in the EMC  are also simulated.

Since the \ks\ decay length is approximately 1/10 the distance traveled by a photon in our time resolution we take all \ks-decay photons as originating at the IP. A prompt photon is defined as a neutral cluster in the EMC, satisfying
the condition $|T-R/c|< {\rm min}(5\sigma_t,\ 2{\rm ns})$, where $T$ is the time of flight (TOF) and $R=|{\bf R}|$
indicates the cluster position with respect to the
detector origin of coordinates. $\sigma_t$ is the total time resolution.
After tagging, we define a signal-enriched sample by requiring two and no more than two prompt photons in the event.
While the minimum energy of photons from \ks\to\gam\gam\ is 197 MeV, photons from \ks\to2\pio\to4\gam\ can be much softer, $E_{\gam}>\,$15.8 MeV. Also at this momentum our resolution is of \ord40\%;. To maximize \ks\to2\pio\ rejection we therefore consider all clusters with $E>\,$7 MeV, and $|\cos(\theta)|<0.93$.
The distribution of photons from \ks\to2\pio\ not detected by the EMC is peaked at $|\cos\theta|\,$=1,
as shown by the MC spectrum in Fig.~\ref{QCALMC}.
\FIGURE{\parbox{10cm}{\figbc qcal-lost;10;}
\caption{Angular distribution of photons from \ks\to2\pio\ with two photons in the EMC. Reconstructed photons solid-line histogram, undetected photons points.}
\label{QCALMC}}

After these cuts, we are left with 550,000 events, a signal  efficiency of \ab83\% and a signal over background
ratio S/B\ab1/300.  The background is mostly from \ks\to2\pio\  events (99.1\%) and a 0.7\% contamination of false $K_L$-crash from $K^+ K^-$
events. There is also a residual background from \ks-decays other than 2\pio: 0.2\% from \pic, 0.02\% from $\pi\ell\nu$.
To improve background rejection, we veto events with photons absorbed by the QCAL.
Fig.~\ref{QCALT} shows the distribution of the difference between the reconstructed and the expected
time of the QCAL signals, $\Delta T_{Q}$.
The in-time peak is due to \ks\to2\pio\  with photons reaching the QCAL.
\FIGURE{\figb dtqcal;7.7;\caption{Inclusive distribution of the difference between the measured
arrival time and the expected time of flight of hits in QCAL for events tagged by a $K_L$-crash with two prompt photons
(solid line) or with a reconstructed $K_S \to \pi^+\pi^-$ decay (points).}
\label{QCALT}}
The oscillating distribution is due to machine background events and shows the period of the beam bunches.
All events having at least one hit in QCAL with energy above threshold and in a time window, TW, defined by
$|\Delta T_{\rm Q}| < 5$~ns are vetoed. This veto removes  $\sim 70$\% of
the background, while retaining high efficiency for the signal. The signal loss is \ab0.04\%.

We must however correct for the signal loss due to the accidental
coincidence with machine background signals in the TW.
The correction is $C_{\rm Q} = 1-P^{\rm TW}_{\rm Q}$,
where $P^{\rm TW}_{\rm Q}$ is the probability of a random coincidence
in the TW. The latter is taken as the average of values obtained in two different out-of-time
windows, one early and one late with respect to the collision time.
We estimate the systematic error from the value of $P^{\rm TW}_{\rm Q}$ obtained from
reconstructed \ks\to\pic\ decays where
no photons are present. We find:
$P^{\rm TW}_{\rm Q}=(3.51 \pm 0.04_{{\rm stat}} \pm 0.26_{{\rm syst}}) \%$.
At the end of the acceptance and QCAL veto selection, we remain with
157 $\times 10^3$ events. The S/B ratio is \ab1/80 at this stage.

\subsection{Kinematic fitting and  event counting}
%
To improve the S/B ratio, we perform a kinematic fit imposing seven constraints: energy and momentum conservation, the kaon mass and the two photon velocities.
Input variables to the fit are the IP coordinates, the \ks\ decay point, the \ks\ momentum $|{\bf p}|$, the interaction points of the two photons in the EMC and the two cluster energies. All of these 15 variables are adjusted by the fit. There is no unmeasured variable to be determined. So this is a 7-C fit with the number of degrees of freedom being dof=7. Fig. 3a, 3b and 6a, show a peak in $\chi^2$ at \ab5 as expected for dof=7.

Fig.~\ref{chi2} shows the $\chi^2$ distribution from the fit for data and MC events, after acceptance selection,
before and after  applying the  QCAL veto.
\FIGURE{\figb chi2-a-b;12;\caption{$\chi^2$ distributions for tagged $K_S$ events with
two prompt photons: before (a) and after (b) QCAL veto.}\label{chi2}}
The background has high $\chi^2$ values.
Rejecting events with $\chi^2>20$ we retain $\sim$ 63\%
of the signal while considerably reducing the background.
The S/B ratio improves from 1/80  to 1/3. After this  cut,
the background is entirely due to $K_S\to2\pio$
events with two undetected photons.
Background, Fig. \ref{scatter}, can be further reduced
using the \ggam\ invariant mass
$M_{\ggam}$, and the photon opening angle in
the kaon rest frame, $\theta^{*}_{\ggam}$.
Since the kinematic fit imposes the kaon mass as a constraint, we use the measured variables values before fitting.
Fig. \ref{scatter} shows plots of $M_{\ggam}$ vs $\cos\theta^{*}_{\ggam}$
for data, MC background and MC signal events.
\FIGURE{\figb scatter_grey;8;\caption{Scatter plot of $M_{\ggam}$ vs $\cos\theta^{*}_{\ggam}$, after pre-selection, for data (a), MC background (b) and MC signal (c). The solid curve
represents the signal dominated region.}
\label{scatter}
}

To check the MC description of the EMC as a function of the photon energy,
we inspect the energy pulls of the kinematic fit for 2\pio\
\ks\ decays. We use a data sample corresponding to $\int\!\L\dt$\ab80 pb\up{-1} and equal MC statistics.
We select tagged \ks-mesons and ask for four prompt photons.
An energy scale correction of $\sim 1.02$ is required to improve the match between MC simulation and data.
After applying this correction, the MC ability to reproduce signal spectra is tested with a control sample of
\kl\to\break\ggam\ events decaying near the beam pi-pe, with the \kl-meson
tagged by a well reconstructed \ks\to\pic\ decay. The BR for \kl\to2\gam\ is \pt5.74,-4, which together with the lifetime, $\tau_{\kl}$\break=\pt5.08,-8, corresponds to an equivalent BR(\kl\to2\gam) of \pt1.6,-6, per cm of \kl\ path. Thus decays within 30 cm of the IP provide a sample of \kl\to2\gam\ larger than that of \ks\to2\gam\ and with a background level from 2\pio\ decays smaller by three orders of magnitude. The \kl\ vertex position is
     calculated by knowing the $K_L$ flight direction and the time
     of flight of the two photons with a precision of $\sim 1.5 cm$.

Data (MC) corresponding to $\int\!\L\dt\,$=200 (450) pb\up{-1} are used. Events are selected as for  the $\ks\to2\gam$ decays, including the kinematic fit. The background is negligible after requiring $\chi^2<20$. A gaussian fit
to the  $M_{\ggam}$  distributions is shown in Fig.~\ref{energyscale}.
\FIGURE{\figb mgg-dat-mc;13;\caption{\gam\gam\ invariant mass for the $K_L\to\gamma\gamma$
decays near the beam pipe.}\label{energyscale}}
Data and MC energy scales agree to better than 0.2\%, \ab1/5 of the error which is quite satisfactory. 
The resolution agrees to 2\%. The $\chi^2$ and $\cos\theta^*_{\gam\gam}$ distributions of the \kl\ events, Fig.~\ref{kl_costhechi2} a and b, confirm
the simulation results for \ks\to2\gam\ decays.

\FIGURE{\figb chi2_thegg_kl;13;\caption{Distributions of $\chi^2$ (a) and $\cos\theta^*_{\gam\gam}$ (b)
for \kl\to\gam\gam\ decays near the IP. Black points are data, grey histogram is the MC simulation.
The plot of $\chi^2$ has been
 done with preselection cuts on the  two photons and
 a cut on $\cos\theta^*_{\gam\gam}$ below -0.998. The $\cos\theta^*_{\gam\gam}$ distribution 
 required a $\chi^2$ cut at 20. }
\label{kl_costhechi2}}

To obtain the number of \ks\to2\gam\ events, we perform a 2 dimensional binned-maximum-likelihood of the the final sample  distribution in the $M_{\gam\gam}$ and $\cos\theta^{*}_{\gam\gam}$ variables.
The likelihood function uses the MC generated signal and background shapes taking into account data and
MC statistics. The fit gives N$(\gamma \gamma)=711 \pm 35$,
with a $\chi^2/{\rm dof}=854/826$. The fit CL is 24.3\%. 

Projections of data and fit are shown in Fig.~\ref{2dresult}.
\FIGURE{\parbox{12cm}{\figbc fit-result;9;}\caption{ Distributions of$\cos(\theta^{*}_{\gam\gam})$ (a)
 and $M_{\gam\gam}$ (b) for the final sample.}\label{2dresult}} The signal  $\cos\theta^{*}_{\ggam}$ distribution
is peaked at $\cos\theta\,$=\minus1 while the $M_{\ggam}$ distribution is gaussian at the $\ks$ mass. The background is less peaked at $\cos\theta\,$=\minus1 and lower and broader in mass.
As an independent  check of the fit quality, we show in
Fig.~\ref{control_plots}.a the $\chi^2$ distribution
for data  and MC after minimization.
\FIGURE{\epsfig{width=11cm,file=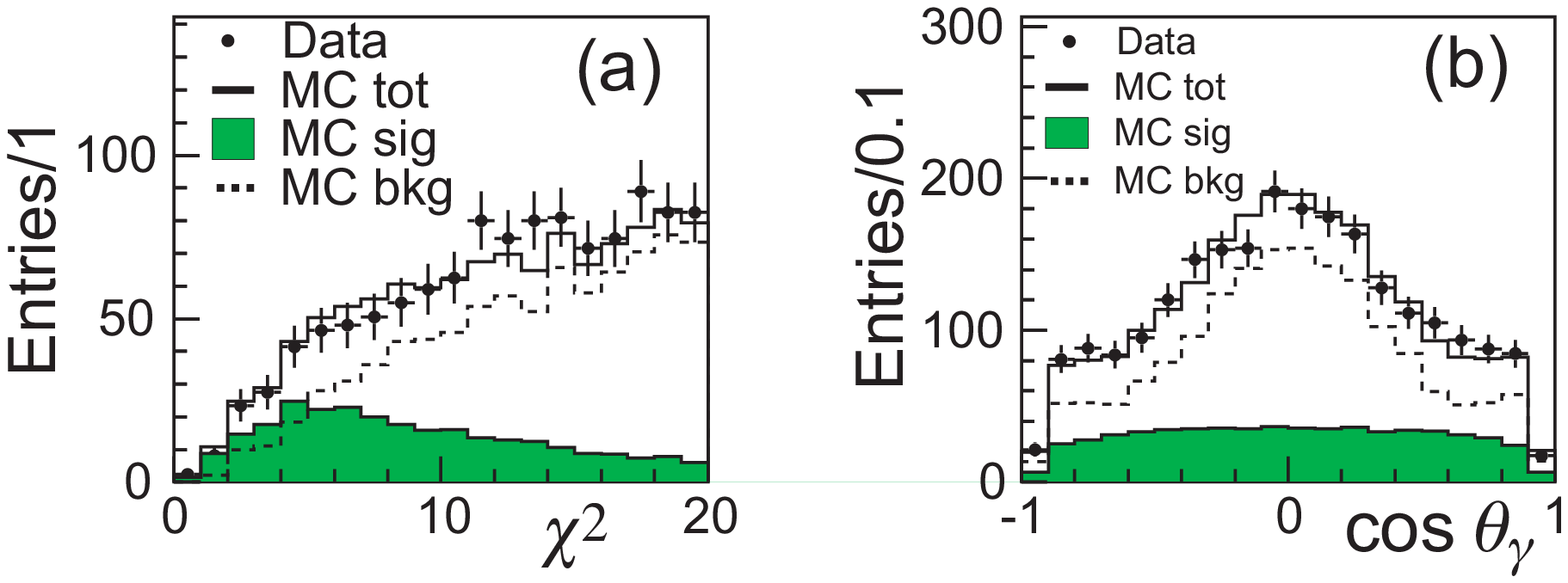}
\caption{Distributions of
$\chi^2$ (a) and inclusive $\cos\theta_{\gamma}$
of the two photons in the event (b) for the final sample.}
\label{control_plots}}
A similar comparison is done also for the angular
photon spectrum (Fig.~\ref{control_plots}.b), which clearly
indicates the presence of a flat component due to signal, as
expected for the two body decay of a spin 0 particle.

\section{Branching ratio evaluation and systematics}
The branching ratio is obtained from $N(\ks\to\gamma\gamma)$ using for normalization the yield for \ks\to2\pio\ in the same sample of tagged \ks-mesons by counting events with four prompt photons:
\begin{equation}
{\rm BR}(K_S \to 2\gamma) = \frac{N(\ks\to\ggam)}{N(\ks\to2\pi^0)} \times
\frac{\epsilon_{{\rm TOT}}(2\pi^0|{\rm tag})}{\epsilon_{{\rm TOT}}
(\gamma\gamma|{\rm tag})}
\times {\rm BR}(K_S \to 2\pi^0) \times {\rm R_{\epsilon}}
\end{equation}
The total efficiencies have been evaluated by MC
after $K_L$-crash tag. The signal total efficiency  is the product
of the efficiencies for the acceptance selection, the QCAL cut
and the $\chi^2$ cut:
\begin{equation}
\epsilon_{\rm TOT}(\gamma\gamma) = \epsilon_{\rm sel}(\gamma\gamma)
\times \epsilon_{\rm Q}(\gamma\gamma) \times \epsilon_{\chi^2}(\gamma\gamma).
\end{equation}
For the normalization sample, the
efficiency is related only to the acceptance of four photons.
The ratio, $R_{\epsilon}$, of all other efficiencies
(triggering, Filfo filter and tagging) between
signal and normalization sample should be identically one. From MC we find  $R_{\epsilon} = 1.001 \pm 0.001_{{\rm stat}}$.
The difference from unity is added as contribution to the
systematic error on the BR.

For the signal selection efficiency we find:
\begin{equation}
 \epsilon_{\rm sel}(\gamma\gamma)=
(82.9 \pm 0.2_{{\rm stat}}\pm 0.2_{{\rm syst}}) \%.
\end{equation}
The large selection efficiency is due to the wide angular coverage
of the calorimeter,  the low energy threshold used and the almost
flat angular distribution of the decay products.  The systematic
error assigned to this efficiency has been found by varying the
data-MC correction of the  cluster reconstruction efficiency.
The efficiency for the QCAL cut
is found from MC to be $\epsilon_{Q}^{MC}(2\gamma) \sim$ 99.96\%.
Applying  the correction due to accidental losses
described in sec.~\ref{presel} we obtain:
\begin{equation}
\epsilon_{\rm Q}(2\gamma) = \epsilon_{\rm Q}^{\rm MC}(2\gamma) \times C_{\rm Q}
= (96.45\pm 0.04_{{\rm stat}} \pm 0.26_{{\rm syst}}) \%.
\end{equation}
The MC efficiency of the $\chi^2$ cut
is $\epsilon_{\chi^2} = (63.3 \pm 0.7)\%.$
The systematic error related to the knowledge of
the data\minus MC difference in the $\chi^2$ scale
has been evaluated by using the $K_L \to \gamma \gamma$
control sample. For the chosen $\chi^2$ cut,
we evaluate the data over MC ratio, $R_\chi$, of the
$\chi^2$ cumulative distributions and
we get $(R_\chi-1) =(-0.5 \pm 1.8)\%$. We conservatively
assign the error on $R$ as the contribution of the
$\chi^2$ scale to the systematic error.

\TABLE{\begin{tabular}{|c|c|c|}
\hline
Source & +$\Delta {\rm BR}/{\rm BR}$ (\%) &
-$\Delta {\rm BR}/{\rm BR}$ (\%) \\
\hline
Trigger, Filter, Tag & 0.10 & 0.10 \\
\hline
Signal acceptance & 0.17 & 0.17 \\
QCAL veto & 0.02 & 0.26 \\
$\chi^2$ scale   & 1.80  & 1.80 \\

\hline
Background shape & 1.04  & 0.98\\
QCAL TW change   & 0.53  & 0.49 \\
$\chi^2$ change  & 0.99  & --- \\
MC Energy scale  & ---   & 0.79 \\
2D-Fit binning   & 0.96  & 0.98 \\
\hline
Normalization sample & 0.15 & 0.15 \\
\hline
\hline
Total & 2.56  &  2.48 \\
\hline
\end{tabular}
\caption{Breakdown of the contributions to the total systematic
error for the BR$(K_S \to \gamma\gamma)$.}
\label{TABSYS}}

The systematic uncertainties connected to the signal
counting have been evaluated by repeating the analysis and
the fit in different ways.
The most delicate point is related to the simulation
of the background shape. The MC shows a
good agreement with data for background-enriched samples
obtained by requiring a complementary cut on $\chi^2$,
such as  $30 <\chi^2 < 500$. Moreover,
to test the fit stability in different regions of the
$M_{\gamma\gamma}$ -- $\cos\theta^*_{\gamma\gamma}$ plane,
we have determined how much the result varies when:
(1) reducing the fit-region along the
$\cos\theta^*_{\gamma\gamma}$
axis moving the lower boundaries from 0.999 to 0.9995 
or (2) fitting only in a signal dominated region shown by
the ellipse in Fig.~\ref{scatter}. The maximum
variation of the BR for these tests is reported as
background shape in Tab.~\ref{TABSYS}.

We have also tested the stability of the branching ratio when
modifying the width of the time window used for the QCAL veto
from $\pm 5$~ns to $\pm 4$, $\pm 6$~ns. Similarly,
the cut in $\chi^2$ has been changed from 20 to 10 and 24.
We have then repeated the fit by applying to the MC
an energy-scale correction of +0.4\%,  a factor of two larger
than what measured with the $K_L \to \gamma \gamma$ control
sample. We have also checked that regrouping the bins of
the 2-D plot by factors from 2 to 5 does not modify substantially
the result. For all of these cases, the maximum variation of the BR
obtained is used as systematic error and shown in
Tab.~\ref{TABSYS}. The sum in quadrature of all entries
is used as total systematic error.

For the normalization  we count $K_S \to 2\pi^0$ tagged
events with four prompt photons. An efficiency of
\begin{equation}
\epsilon_{\rm sel}(2\pi^0) = (65.0 \pm 0.2_{{\rm stat}} \pm {0.1}_{{\rm syst}}) \%
\end{equation}
is found by MC. As for the signal, the systematic uncertainty
related to the cluster detection efficiency
is evaluated by varying the data-MC correction curves.
After correcting for $\epsilon_{\rm sel}(2\pi^0)$, a  number of
$(190.5 \pm 0.2) \times 10^6$  $K_S \to 2\pi^0$
tagged events is obtained. The systematic uncertainty
related to the presence of machine background
clusters, fragmentation and merging of clusters
is  estimated by repeating the measurement
in an inclusive way and counting tagged events with 3, 4 and 5
photons. The overall systematic error for the normalization
sample is reported in Tab.~\ref{TABSYS}.

To evaluate BR($K_S\to\gamma\gamma$) we use the latest PDG~\cite{pdg06}
value BR($K_S \to 2 \pi^0$)$=(30.69 \pm 0.05)$\%. See also \cite{klopo}.
We obtain:
\begin{equation}
{\rm BR}(K_S \to \gamma \gamma) = (2.26 \pm
0.12_{\rm stat}\pm 0.06_{\rm syst} )\times 10^{-6}.
\end{equation}
We have repeated  the measurement by subdividing the data in two
sets to check stability for the slightly different running
conditions: 1) 0.4 fb$^{-1}$ from 2001-2002
and 2) 1.5 fb$^{-1}$ for 2004-2005.
Also the simulation has been divided accordingly.
We get
$BR(K_S \to \gamma \gamma)= (2.24 \pm 0.30_{\rm stat})\times 10^{-6}$
in 2001-2002 and
$BR(K_S \to \gamma \gamma)= (2.26 \pm 0.13_{\rm stat})\times 10^{-6}$
in 2004-2005, which are in excellent agreement.

Fig.~\ref{finalres} shows our result and other existing measurements of BR(\ks$\to$\gam\gam) as well as the $\mathcal O(p^4)$ $\chi$PT theoretical prediction. There is a 3 $\sigma$'s discrepancy between the present result and the
measurement of NA48.
\FIGURE{\epsfig{width=6cm,file=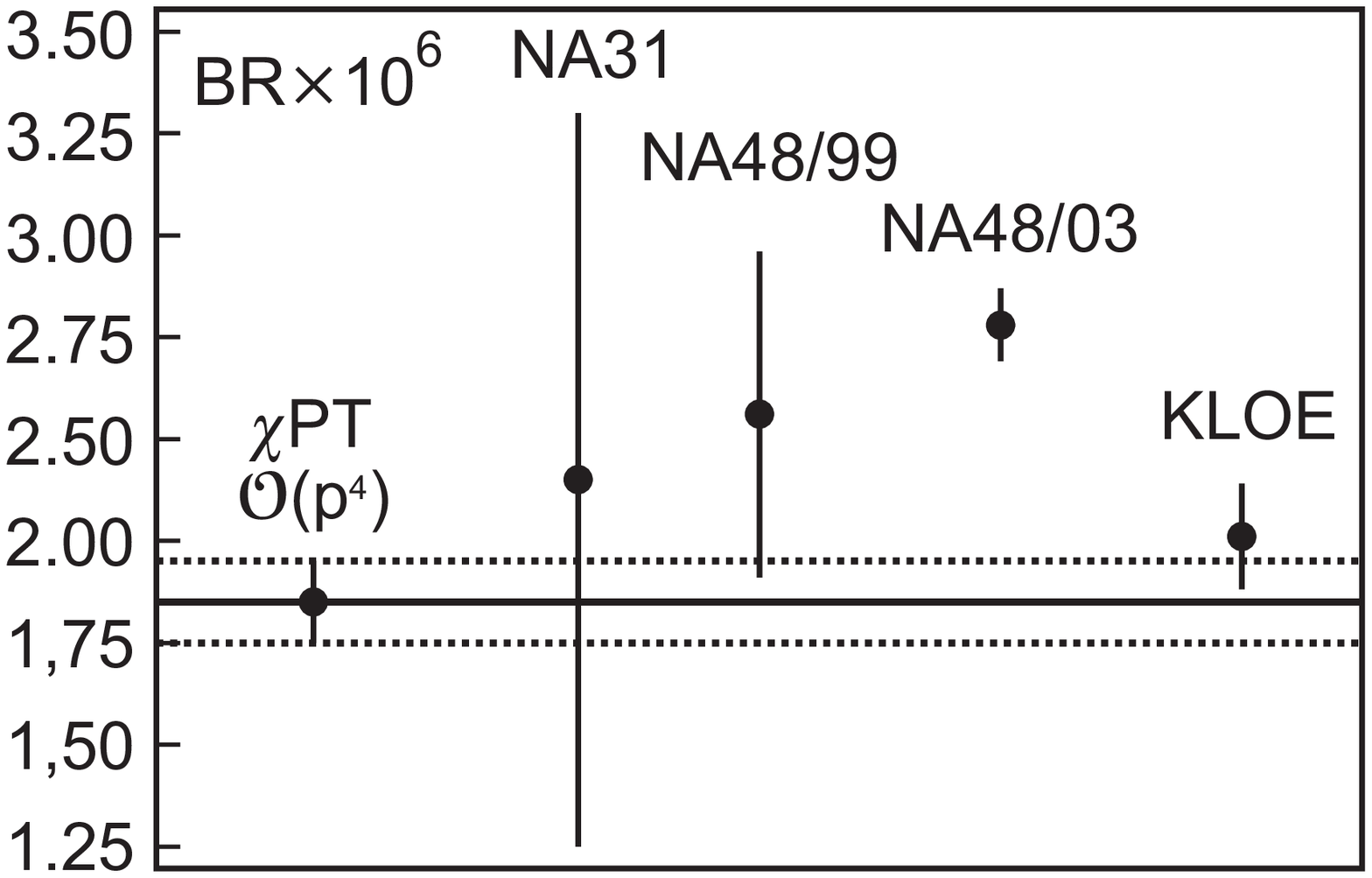}
\caption{Comparison of BR$(K_S \to \gamma \gamma)$ measurements and $\chi$PT predictions.}
\label{finalres}}

\section{Conclusion}
From $\sim$ 2 billion $\phi$ mesons
collected with KLOE at DA$\Phi$NE,
we have measured the ${\rm BR}(K_S\to \gamma\gamma)$
with a 5.3\% statistical  uncertainty and a  $\sim 2$\% systematic error.
We obtain a BR result which deviates by 3 $\sigma$'s  from the previous
best determination. 
Precise $\chi PT$ theory calculation for this decay are done
at $\mathcal O(p^4)$. Higher order effects are predicted
to be at most of the order of $\sim$ 20\% of the
$\mathcal O(p^4)$  decay amplitude. Our measurement is consistent
with negligible higher order corrections.

\section{Acknowledgements}
We thank the DA$\Phi$NE team for their efforts in maintaining
low background running conditions and their collaboration
during all data-taking. We want to thank our technical staff:
G.F. Fortugno and F. Sbor\-zacchi for their dedicated work to ensure an
efficient operation of
the KLOE Computing Center;
M. Anelli for his continuous support to the gas system and the safety of
the
detector;
A. Balla, M. Gatta, G. Corradi and G. Papalino for the maintenance of the
electronics;
M. Santoni, G. Paoluzzi and R. Rosellini for the general support to the
detector;
C. Piscitelli for his help during major maintenance periods.
This work was supported in part
by EURODAPHNE, contract FMRX-CT98-0169;
by the German Federal Ministry of Education and Research (BMBF) contract 06-KA-957;
by the German Research Foundation (DFG),'Emmy Noether Programme',
contracts DE839/1-4; by DOE grant DE-FG-02-97ER41027;
and by the EU Integrated
Infrastructure
Initiative HadronPhysics Project under contract number
RII3-CT-2004-506078.

\catcode`@=11
\catcode`\%=12
\catcode`\|=14
\newcommand\nca[3]    {\@spires{NUCIA
        {{\it Nuovo Cim.\ }{\bf A#1} (#2) #3}}
\renewcommand\jphg[3]   {\@spires{JPHGB
        {{\it J. Phys.\ }{\bf G #1} (#2) #3}}
\catcode`\%=14
\catcode`\|=12
\catcode`@=12

\bibliographystyle{JHEP}

\end{document}